\documentclass[aps,rmp,reprint,superscriptaddress]{revtex4-2}

\usepackage{enumerate,appendix}
\usepackage{commath,braket}
\usepackage{color,calc,graphicx}
\usepackage[usenames,dvipsnames,svgnames,table,cmyk,hyperref]{xcolor}
\usepackage{epstopdf}
\usepackage{makecell}
\usepackage[charter,cal=cmcal,sfscaled=false]{mathdesign}
\definecolor{oeawblue}{cmyk}{0.9,0.68,0,0}
\definecolor{iqoqiblue}{cmyk}{0.76,0.11,0,0}
\usepackage{hyperref}

\hypersetup{
	colorlinks = true,
	urlcolor = {blue},
	citecolor = {magenta},
	linkcolor= {blue}
}
\newcommand{\tr}{{\rm tr}}
\newcommand{\degree}{{}^{\circ}}

\def\id{{\mathbb I}}

\def\bra#1{\langle#1|} \def\ket#1{|#1\rangle}

\newcommand{\be}{\begin{equation}}
\newcommand{\ee}{\end{equation}}

\begin{document}

\title{Testing real quantum theory in an optical quantum network}

\author{Zheng-Da Li}
\affiliation{Shenzhen Institute for Quantum Science and Engineering and Department of Physics, Southern University of Science and Technology, Shenzhen, 518055, China}
\affiliation{Guangdong Provincial Key Laboratory of Quantum Science and Engineering, Southern University of Science and Technology, Shenzhen, 518055, China}

\author{ Ya-Li Mao}
\affiliation{Shenzhen Institute for Quantum Science and Engineering and Department of Physics, Southern University of Science and Technology, Shenzhen, 518055, China}
\affiliation{Guangdong Provincial Key Laboratory of Quantum Science and Engineering, Southern University of Science and Technology, Shenzhen, 518055, China}

\author{Mirjam Weilenmann}
\affiliation{Institute for Quantum Optics and Quantum Information - IQOQI Vienna, Austrian Academy of Sciences, Boltzmanngasse 3, 1090 Vienna, Austria}

\author{Armin Tavakoli}
\affiliation{Institute for Quantum Optics and Quantum Information - IQOQI Vienna, Austrian Academy of Sciences, Boltzmanngasse 3, 1090 Vienna, Austria}
\affiliation{Institute for Atomic and Subatomic Physics, Vienna University of Technology, 1020 Vienna, Austria}

\author{Hu Chen}
\affiliation{Shenzhen Institute for Quantum Science and Engineering and Department of Physics, Southern University of Science and Technology, Shenzhen, 518055, China}
\affiliation{Guangdong Provincial Key Laboratory of Quantum Science and Engineering, Southern University of Science and Technology, Shenzhen, 518055, China}

\author{Lixin Feng}
\affiliation{Shenzhen Institute for Quantum Science and Engineering and Department of Physics, Southern University of Science and Technology, Shenzhen, 518055, China}
\affiliation{Guangdong Provincial Key Laboratory of Quantum Science and Engineering, Southern University of Science and Technology, Shenzhen, 518055, China}

\author{Sheng-Jun Yang}
\affiliation{Shenzhen Institute for Quantum Science and Engineering and Department of Physics, Southern University of Science and Technology, Shenzhen, 518055, China}
\affiliation{Guangdong Provincial Key Laboratory of Quantum Science and Engineering, Southern University of Science and Technology, Shenzhen, 518055, China}

\author{Marc-Olivier Renou}
\affiliation{ICFO-Institut de Ciencies Fotoniques, The Barcelona Institute of Science and Technology, 08860 Castelldefels (Barcelona), Spain}

\author{David Trillo}
\affiliation{Institute for Quantum Optics and Quantum Information - IQOQI Vienna, Austrian Academy of Sciences, Boltzmanngasse 3, 1090 Vienna, Austria}

\author{Thinh P. Le}
\affiliation{Institute for Quantum Optics and Quantum Information - IQOQI Vienna, Austrian Academy of Sciences, Boltzmanngasse 3, 1090 Vienna, Austria}

\author{Nicolas Gisin}
\affiliation{Group of Applied Physics, University of Geneva, 1211 Geneva 4, Switzerland}
\affiliation{Schaffhausen Institute of Technology - SIT, Geneva, Switzerland}

\author{Antonio Ac\'in}
\affiliation{ICFO-Institut de Ciencies Fotoniques, The Barcelona Institute of Science and Technology, 08860 Castelldefels (Barcelona), Spain}
\affiliation{ICREA, Pg. Lluis Companys 23, 08010 Barcelona, Spain}

\author{Miguel Navascu\'es}
\affiliation{Institute for Quantum Optics and Quantum Information - IQOQI Vienna, Austrian Academy of Sciences, Boltzmanngasse 3, 1090 Vienna, Austria}

\author{Zizhu Wang}
\affiliation{Institute of Fundamental and Frontier Sciences, University of Electronic Science and Technology of China, Chengdu 610054, China}

\author{Jingyun Fan}
\affiliation{Shenzhen Institute for Quantum Science and Engineering and Department of Physics, Southern University of Science and Technology, Shenzhen, 518055, China}
\affiliation{Guangdong Provincial Key Laboratory of Quantum Science and Engineering, Southern University of Science and Technology, Shenzhen, 518055, China}

\begin{abstract}
Quantum theory is commonly formulated in complex Hilbert spaces. However, the  question of whether complex numbers need to be given a fundamental role in the theory has been debated since its  pioneering days. Recently it has been shown that tests in the spirit of a Bell inequality can reveal quantum predictions in entanglement swapping scenarios that cannot be modelled by the natural real-number analog of standard quantum theory.
Here, we tailor such tests for implementation in state-of-the-art photonic systems. We experimentally demonstrate quantum correlations in a network of three parties and two independent EPR sources that violate the constraints of real quantum theory by over $4.5$ standard deviations, hence disproving real quantum theory as a universal physical theory. 
\end{abstract}

\maketitle

The original formulation of quantum theory postulates states and measurements as operators in complex Hilbert spaces and uses tensor products to model system composition~\cite{DiracBook,vonNeumannBook}.
However, already some pioneers of the theory favored a real quantum theory over a complex quantum theory, i.e.~using only real numbers in its mathematical formulation~\cite{QuantumLetters2011}. 
The general debate on the  role of complex numbers in quantum theory has continued into the present~\cite{Stueckelberg1960, Wootters1981, Wootters1990, HardyWootters2012, Aleksandrova2013,Moretti2017,Drechsel2019,GuoImaginary1}.

Another long-standing debate in the foundations of quantum theory, nowadays settled, concerned the existence of local hidden variable theories to describe our world. Bell pioneered the idea of studying correlations in the outcome statistics of experiments to infer fundamental properties of their underlying physics~\cite{Bell1964}. In recent years, experimental implementations of such Bell tests have successfully ruled out local hidden variable theories  \cite{HansonLoopholeFree2015,ZeilingerLoopholeFree2015,ShalmLoopholefree,Rosenfeld2017,LiPRL2018}. Surprisingly, it was further shown that a natural generalization of Bell's test in a network can, contrary to their traditional counterparts \cite{Pal2008, McKague2009}, distinguish complex quantum theory from real quantum theory~\cite{Renou2021}. In a network in which parties are connected through several independent sources \cite{Branciard2010, Fritz2012, Tavakoli2021}, real quantum theory does not agree with all predictions of complex quantum theory \cite{Renou2021}. This paves the way for experimentally distinguishing between the two theories in a quantum network based on independent sources. Here and in the rest of this paper, \emph{real quantum theory} refers to a theory in which the real Hilbert spaces of independent systems are combined by the tensor product.

\begin{figure}[htbp!]
	\centering
	\includegraphics[width=0.5\textwidth]{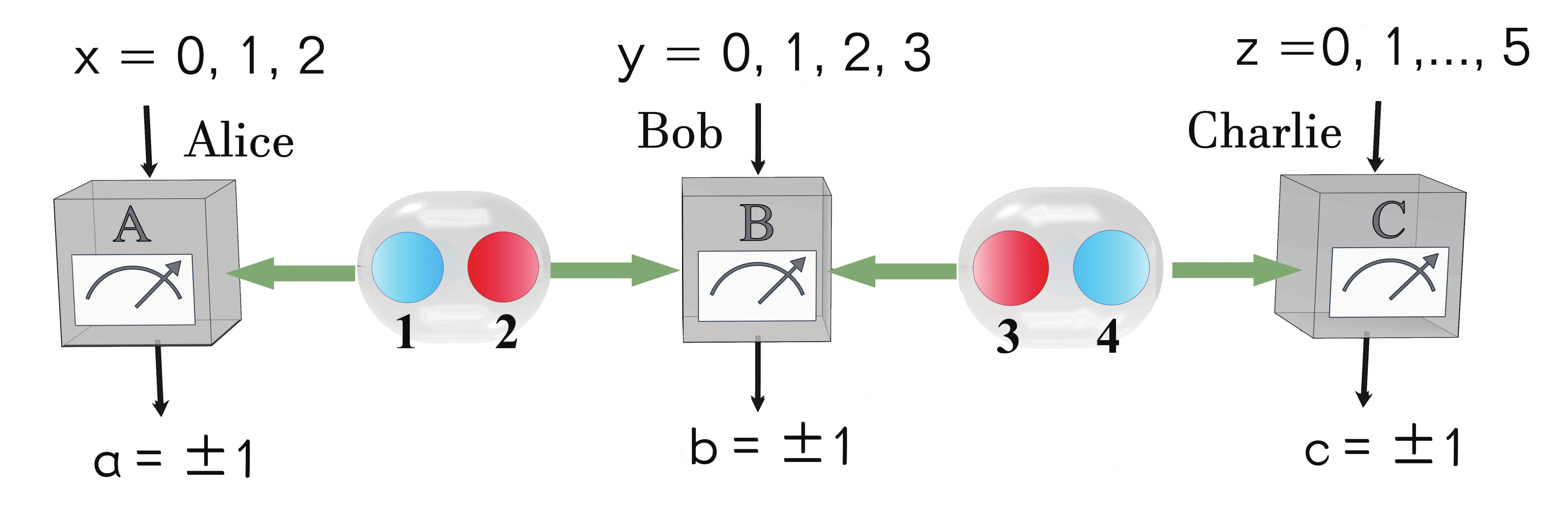}
	\caption{The entanglement swapping scenario. Two sources, that may at most be classically correlated, distribute an entangled pair between Alice and Bob, and Bob and Charlie respectively. Alice, Bob and Charlie each select one of three, four and six inputs respectively and perform  corresponding quantum measurements on their respective shares. Each measurement gives one of two possible outcomes.}
	\label{fig:CP-f1}
\end{figure}

For this purpose, the simplest quantum network, describing the entanglement swapping scenario, suffices. As shown in Fig. \ref{fig:CP-f1}, two independent sources each emit an EPR pair, the first one shared between Alice and Bob, and the second shared between Bob and Charlie. If Bob projects both of his particles onto an entangled basis, then Alice and Charlie are left in an entangled state when conditioned on Bob's outcome~\cite{Zukowski1993}. By appropriately combining the input-output probabilities of the network we arrive at a Bell-type correlation function, whose maximum value in real quantum theory  can be upper bounded. Any experimental violation of this upper bound would disprove the universal validity of real quantum theory~\cite{Renou2021}. 

Photons emerge as a natural platform for realizing quantum networks of scalable distance as they are suitable carriers of quantum information. Therefore it is natural to use a photonic quantum network to test real quantum theory.  Moreover, the pivotal requirement of independent sources can be stringently met in photonic implementations, whereas a recent similar experiment based on superconducting qubits implemented both sources on the same  chip~\cite{Chen2021}. However, photonic implementations come with the difficulty that the protocol proposed in \cite{Renou2021} requires a complete Bell state measurement. Even though such a measurement can be implemented using superconducting qubits~\cite{Chen2021}, it is not amenable to the tools of standard linear optics~\cite{Lutkenhaus1999, Calsamiglia2001}, unless additional degrees of freedom are introduced and controlled~\cite{Schuck2006}. Here, we resolve this issue by developing a new  protocol that uses a partial Bell state measurement~\cite{Weinfurter1994, Braunstein1995}. 
 
Consider the scenario illustrated in Fig. \ref{fig:CP-f1}. We focus on a quantum protocol in which each source emits the EPR state $\ket{\Phi^+}=\frac{\ket{00}+\ket{11}}{\sqrt{2}}$. Alice, Bob and Charlie independently perform measurements with random inputs $x\in\{0,1,2\}$, $y\in\{0,1,2,3\}$ and $z\in\{0,1,2,3,4,5\}$, respectively. Their measurement outcomes are labeled $a,b,c\in\{+1,-1\}$. Alice's three measurements are chosen as $\{\sigma_X, \sigma_Y, \sigma_Z\}$ and Charlie's six measurements are chosen as $\{\frac{(\sigma_X+ \sigma_Y)}{\sqrt{2}},  \frac{(\sigma_X- \sigma_Y)}{\sqrt{2}}, \frac{(\sigma_Y + \sigma_Z)}{\sqrt{2}}, \frac{(\sigma_Y - \sigma_Z)}{\sqrt{2}}, \frac{(\sigma_X+ \sigma_Z)}{\sqrt{2}},\frac{(\sigma_X- \sigma_Z)}{\sqrt{2}}\}$, where $\sigma_X, \sigma_Y, \sigma_Z$ are Pauli observables. 
Bob's four measurements each correspond to discriminating one of the four Bell states. Specifically, the outcome $b=1$  corresponds respectively to a projection onto the Bell state $\ket{\Phi^+}$, 
 $\ket{\Phi^-}=\frac{\ket{00}-\ket{11}}{\sqrt{2}}$, $\ket{\Psi^-}=\frac{\ket{01}-\ket{10}}{\sqrt{2}}$ and $\ket{\Psi^+}=\frac{\ket{01}+\ket{10}}{\sqrt{2}}$.
As in any Bell test, by suitably combining the probabilities  $p(a,b,c|x,y,z)$, we define a Bell-type correlation function,
\begin{equation}
W=\frac{1}{5}\sum_{y=0}^{3}T_y-\frac{4}{5}\sum_{y=0}^{3}p(b=1|y), \label{Wfunction}
\end{equation}
where $y=y_2y_1\in\{00,01,10,11\}$ in binary notation and
$	T_y=(-1)^{y_1+y_2}(S_{11y}+S_{12y})-(-1)^{y_1}(S_{21y}-S_{22y})
	+(-1)^{y_1+y_2}(S_{15y}+S_{16y})+(-1)^{y_2}(S_{35y}-S_{36y})
	-(-1)^{y_1}(S_{23y}+S_{24y})+(-1)^{y_2}(S_{33y}-S_{34y}),
$ 
with $S_{xzy}=\sum_{a,c=\pm1} ac p(a,b=1,c|x,y,z)$.  Using the tools developed in~\cite{Renou2021}, the value of~(\ref{Wfunction}) in real quantum theory is upper bounded by (Supplementary)
\begin{equation}\label{RQT}
	W_\text{RQT}\lesssim 0.7486.
\end{equation}
This bound holds even if Alice, Bob and Charlie are allowed global shared classical randomness, i.e.~the two sources may be classically, but not quantumly correlated. In turn, complex quantum theory predicts the value
\begin{equation}
	W_\text{CQT}=\frac{6\sqrt{2}-4}{5}\approx 0.8971.
\end{equation}
 Hence an experimental observation of $W_\text{EXP}>0.7486$ is sufficient to rule out real quantum theory.

\begin{figure*}
	\centering
	\includegraphics[width=\linewidth]{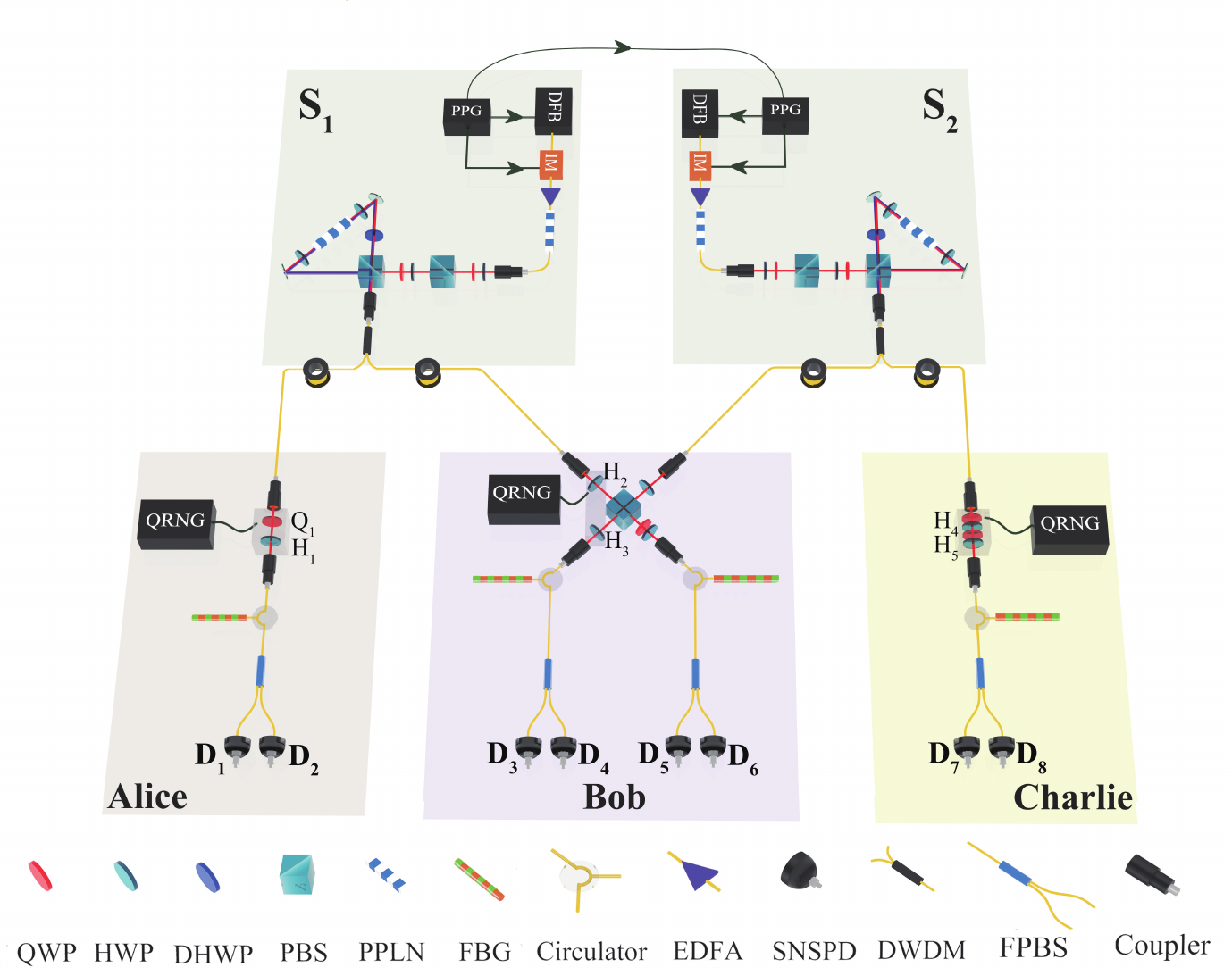}
	\caption{Schematic of the experiment. The setup consists of two EPR sources, $S_1$ and $S_2$, and three measurement nodes, Alice, Bob, and Charlie. In each EPR source, the laser pulse is injected into a Sagnac loop interferometer containing a PPLN crystal to produce a pair of photons in Bell state $|\Phi^+\rangle$ via the SPDC process~\cite{Sun2019}. The EPR source delivers the signal photon to Bob and the idler photon to Alice (Charlie). Bob performs measurement with the two signal photons, which prepares Alice and Charlie in a Bell state. Alice and Charlie measure the idler photons according to the inputs from the quantum random number generator (QRNG). PPG: pulse pattern generator, PPG in EPR source $S_1$ triggers the PPG in EPR source $S_2$, DFB: distributed feedback laser, IM: intensity modulator, PBS: polarization beam splitter, HWP: half-wave plates, QWP: quarter wave plate, DHWP: HWP for dual wavelength, SPBS: spatial PBS, FBG: fibre bragg grating, $D_1, D_2, D_3, D_4,D_5, D_6,D_7, D_8$ are superconducting nanowire single-photon detector (SNSPD), FPBS: fibre PBS, DWDM: dense wavelength division multiplexer.
	}
	\label{fig:CP-f2}
\end{figure*}

A schematic of our optical quantum network implementing the states and measurements above is depicted in Fig.~\ref{fig:CP-f2}. By driving a Type-0 spontaneous parametric down-conversion (SPDC) process in a periodically poled MgO doped Lithium Niobate (PPLN) crystal with the pump laser at $\lambda_p=779$ nm~\cite{Sun2019}, each EPR source probabilistically emits a pair of photons in state $\ket{\Phi^+}=\frac{\ket{HH}+\ket{VV}}{\sqrt{2}}$ at the phase-matched wavelengths 1560 nm (signal) and 1556 nm (idler), where $\ket{H}$ and $\ket{V}$ represent respectively the horizontal and vertical polarization states.  
The two EPR sources each deliver their signal photon to Bob and their idler photon to Alice and Charlie, respectively. To realize the protocol measurements, Bob lets the two signal photons sequentially pass polarization beamsplitters (PBS) and directs the four outputs to single-photon detectors ($D_3, D_4, D_5, D_6$) via optical fibers. The half-wave plates (HWPs) before and after the first PBS are adjusted to the proper orientations upon receiving $y$.
The resulting two-photon coincidence detection between $D_3$ and $D_5$ or $D_4$ and $D_6$ assigns $b=+1$ and prepares Alice and Charlie to be in the Bell state $|\Phi^+\rangle, |\Phi^- \rangle, |\Psi^- \rangle$ , or $|\Psi^+ \rangle$ according to the random input $y \in \{0,1,2,3\}$, respectively. 
Similarly, upon receiving the random inputs $x$ and $z$, Alice and Charlie each respectively adjust the relevant HWP and quarter-wave plate (QWP) to perform the single-qubit measurements outlined in the above. 
The single-photon detection events at each detector are time-tagged to produce the correlation analysis (Supplementary, Tables 1-2). 
\begin{figure*}
	\centering
	\includegraphics[width=\linewidth]{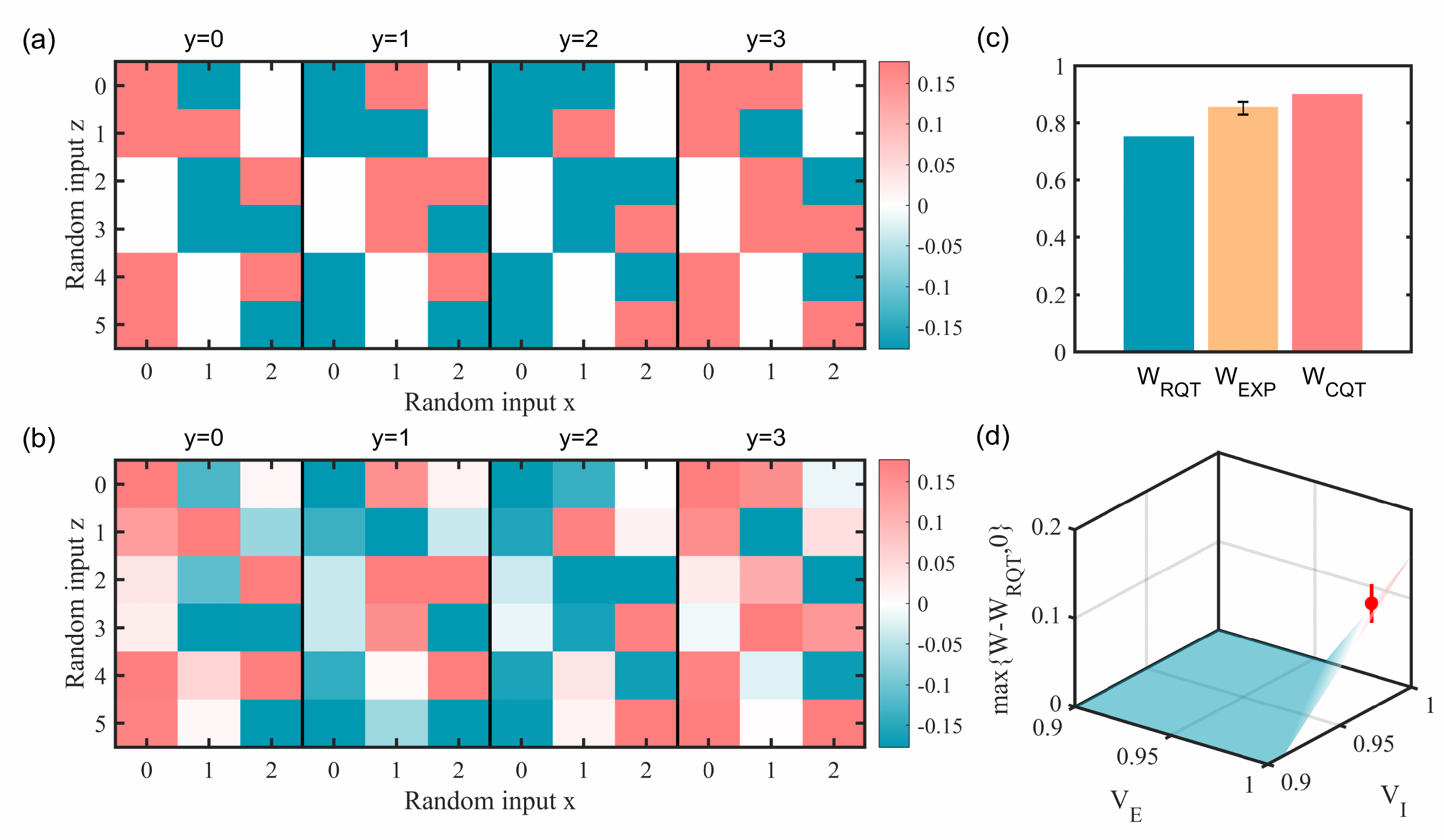}
	\caption{Experimental results.The ideal (a) and experimentally measured (b) values $S_{xzy}=\Sigma_{a,c=\pm1}acp(a,b=1,c|x,y,z)$.  ($\bf{C}$) The values of $W$ in different scenarios. $W_\text{RQT}$: real quantum theory, $W_\text{SQT}$: complex quantum theory, $W_\text{EXP}$: experimental result in our optical quantum network. ($\bf{D}$) An emulation of the experiment based on complex quantum theory. The value of red dot is given by $W_\text{EXP}-W_\text{RQT}$. Error bars shown in (c) and (d) represent one standard deviation in the experiment.	}
	\label{fig:CP-f3}
\end{figure*}

In order to falsify real quantum theory, it is essential to stringently meet the requirement that the two EPR sources are independent up to classical synchronization~\cite{Renou2021}. 
In our network experiment, the time reference of all events is set to the 15 GHz internal clock of a pulse pattern generator (PPG) in the EPR source $S_1$, which triggers the PPG in the EPR source $S_2$. In each EPR source, the PPG sends triggers at a rate of 250 MHz to enable the distributed feedback (DFB) laser to switch on the electric current. Switching from below to above the lasing threshold, the DFB laser emits 2 ns laser pulse at the wavelength of 1558 nm with a randomized phase per trigger~\cite{Sun2019}. We further shorten the pulse width to 90 ps with an intensity modulator (IM). After passing through an erbium-doped-fibre-amplifier (EDFA), a PPLN waveguide to double the frequency, and a wavelength-division multiplex filter, the produced laser pulses at $\lambda_p=779$ nm drive the SPDC process as described in the above, for which we keep the photon-pair production rate at about 0.0025 per trigger to strongly mitigate the multi-photon effect.  We pass photons through fiber Bragg gratings (FBGs) with bandwidths of 3.3 GHz before entering single-photon detectors to suppress the spectral distinguishability between photons from different EPR sources. Quantum tomography measurements indicate that the state fidelity is greater than $0.99$ for the two-photon states produced at EPR sources $S_1$ and $S_2$ with respect to the targeted Bell state $|\Phi^+\rangle$ and greater than $0.96$ for the two-photon state of Alice and Charlie with respect to the Bell states $|\Phi^+\rangle$,  $|\Phi^-\rangle$,  $|\Psi^+\rangle$, and $|\Psi^-\rangle$ after entanglement swapping, respectively. 
The lack of coherence between pulses of the same laser and of different lasers ensures that the independence of the sources is stringently enforced in the network experiment.

We switch the measurement settings every 30 seconds, reserving the first 10 seconds to reset the measurement settings, including quantum random number generation and wave-plate rotation, and the remaining 20 seconds for data collection. The average four-photon coincident detection event rate is 1.24 per switching cycle. We collect 26954 four-photon coincidence events in 21742 switching cycles. Since we only record Bob's outcome $b=+1$, we use our characterisation of the efficiency of the detection device to estimate the associated probability. For each cycle, we estimate $p(b=+1|y)$ through the quantity $\frac{N_{b=+1|y}N_AN_C}{N_{AC}N_{AB}N_{BC}}$, where $N_{AC}$, $N_{AB}$, $N_{BC}$ are the two-party photon coincidence numbers, $N_A$ and $N_C$ are the one-party photon detection numbers and  $N_{b=+1|y}$ is the number of four-photon coincidence events when Bob obtains outcome $b = +1$ (Supplementary).
In Figs.~\ref{fig:CP-f3}(a) and (b), we compare the $3\times 4 \times 6 =72$ different experimentally measured values of $S_{xzy}$, averaged with respect to different cycles, with their corresponding theoretical values. The results uphold a good agreement. Putting all together, we then obtain $W_{\text{EXP}}=0.8508\pm0.0218$, which exceeds the upper bound $W_{\text{RQT}}\approx0.7486$ set by real  quantum theory by $4.70$ standard deviations (Fig.~\ref{fig:CP-f3}(c)). 
We also perform an emulation of the experiment based on complex quantum theory. Consider that the EPR source emits a non-ideal state, $\rho_{EPR}=v_{E}\ket{\Phi^+}\bra{\Phi^+}+(1-v_{E})I/4$, and the photons from the two EPR sources interfere with a non-ideal visibility $v_I$. With $v_E=0.9909$ and $v_{I}=0.9844$ determined in the experiment,  complex quantum theory predicts $W=0.8404$, which is consistent with the experimental result $W_{\text{EXP}}$ within a standard deviation (Fig.~\ref{fig:CP-f3}(d)).

We have experimentally demonstrated that real quantum theory is incompatible with the observed data in our optical quantum network experiment.
The independent sources in the network guarantee that the observed correlations can not be simulated by real quantum theory~\cite{Renou2021}. The rapid development in photonic or hybrid quantum technologies, in particular with regard to more efficient detectors, faster switching, higher-quality entanglement sources and longer-distance entanglement distribution, leaves us optimistic of even more sophisticated future experiments. Research in quantum networks has enjoyed a rapid growth thanks to the role they are likely to play in quantum communications, distributed quantum computing and the future quantum internet. In addition to these potential applications, our work highlights that they are also powerful frameworks for devising and implementing tests of fundamental aspects of quantum theory. 

D.T. is a recipient of a DOC Fellowship of the Austrian Academy of Sciences at the Institute of Quantum Optics and Quantum Information (IQOQI), Vienna. T.P.L. and M.W. are supported by the Lise Meitner Fellowship of the Austrian Academy of Sciences (project numbers M 2812-N and M 3109-N respectively).
Z.-D.L, Y.-L.M., S.-J.Y., and J.F. are supported by the Key-Area Research and Development Program of Guangdong Province Grant No.2020B0303010001, Grant No.2019ZT08X324, and Guangdong Provincial Key Laboratory Grant No.2019B121203002.
Z.W. is supported by the National Key R\&D Program of China (No. 2021YFE0113100, No. 2018YFA0306703) and Sichuan Innovative Research Team Support Fund (2021JDTD0028).
 M.-O.R. and A.T. are supported by the Swiss National Fund Early
Mobility Grants P2GEP2 191444 and P2GEP2 194800 respectively. M.-O.R and A.A acknowledge support from the Government of Spain (FIS2020-TRANQI and Severo Ochoa CEX2019-000910-S), Fundacio Cellex, Fundacio Mir-Puig, Generalitat de Catalunya (CERCA, AGAUR SGR 1381 and QuantumCAT), the ERC AdG CERQUTE, the AXA Chair in Quantum Information Science. N.G. is supported by the Swiss NCCR SwissMap.

\bibliography{ReallyComplexBiblio}

\clearpage

\begin{appendix}
\setcounter{figure}{0}
\renewcommand{\thefigure}{S\arabic{figure}}

\section{Derivation of inequality (\ref{RQT}) in main text}
\label{AppendixTheory}

In the entanglement swapping scenario (see Fig. 1), the two independent sources can share any real quantum states $\rho_{AB}$ and $\rho_{B'C}$ between Alice and Bob and Bob and Charlie, respectively. Alice, Bob and Charlie then perform measurements (with outcomes $a$, $b$ and $c$) on their respective shares, which are denoted as $\{E_{a|x}:a\}$, $\{F_{b|z}:b\}$ and $\{G_{c|z}:c\}$. Mathematically, these are complete sets of projectors for every input $x$, $y$ and $z$. This means that the input-output correlations observed by the parties are
\begin{equation}
P(a,b,c|x,y,z)=\tr\left((\rho_{AB}\otimes\rho_{B'C})(E_{a|x}\otimes F_{b|z}\otimes G_{c|z})\right).
\end{equation}
Notice that the measurements $\{F_{b|z}:b\}$ act on both subsystems held by Bob ($B$ and $B'$).

To prove Inequality (\ref{RQT}) in main text, we adapt the semidefinite-programming (SDP) \cite{sdp} relaxation given in \cite{Renou2021,Moroder2013, Navascues2007} to study a related Bell-type inequality where Bob just had a single measurement setting, see the original paper for more details. 

Let $\rho^{AC}_{b|y}$ be the non-normalized state in Alice and Charlie's systems when Bob conducts measurement $y$ and obtains outcome $b$. Consider now the local completely positive maps given by

\begin{align}
	&\Omega_A(\eta)=\sum_{\alpha,\alpha'}\tr(E^\dagger_{\alpha}\eta E_{\alpha'})\ket{\alpha}\bra{\alpha'},\nonumber\\
	&\Omega_C(\eta)=\sum_{\gamma,\gamma'}\tr(G^\dagger_{\gamma}\eta G_{\gamma'})\ket{\gamma}\bra{\gamma'},
\end{align}
\noindent where $E_{\alpha}, G_{\gamma}$ respectively denote monomials of the operators $\{E_{a|x}\}, \{G_{c|z}\}$, and the sums are over all monomials of degree smaller than or equal to a given order $N$ (in that sum, we include the identity operator, which we regard as a monomial of degree $0$). 

We next define the matrices $\Gamma_{b|y}\equiv \Omega_A\otimes\Omega_C(\rho^{AC}_{b|y})$. From the definition, it is clear that these matrices must be positive semidefinite, and, as observed in \cite{Renou2021}, some of their entries correspond to the observed measurement probabilities $P(a,b,c|x,y,z)$. Moreover, some of the remaining entries must be identical to reflect the fact that products of different pairs of operators can result in the same operator (e.g.: $E_{0|0}E_{0|1}=E_{0|0}(E_{0|1})^\dagger=\id\cdot(E_{0|1}E_{0|0})^\dagger$). In addition, the quantity

\begin{equation}
	\Gamma:=\sum_b \Gamma_{b|y}
\end{equation}
\noindent must be independent of $y$, since it represents the action of $\Omega_A\otimes\Omega_C$ over Alice and Charlie's joint state $\sum_b\rho^{AC}_{b|y}=\rho^{AC}$ before Bob carried out his measurement. Furthermore, since $\rho^{AC}$ is a real-separable state (namely, a convex combination of real product states), so must be the non-normalized bipartite state $\Gamma$. In particular, $\Gamma$ must be invariant under partial transposition of Alice's system \cite{real_quantum2}. 

Let $H^N_{a,c|x,z}$ be the matrix such that $\tr(\Gamma_{b|y} H^N_{a,c|x,z})=P(a,b,c|x,y,z)$, and let $S_N$ be the matrix subspace where $\{\Gamma^y_b\}_{b,y}$ live (it is non-trivial on account of the aforementioned identities between some of the matrix entries). Summing it all up, we have that the maximum value of (\ref{RQT}) in main text in real quantum theory is upper bounded by the following semidefinite program:

\begin{align}
	&\max_{\vec{\Gamma}, P}  W(P),\nonumber\\
	\mbox{s. t. }  &\Gamma_{b|y}\in S_N,\Gamma_{b|y}\geq 0,\forall b,y,\nonumber\\
	&\Gamma^{T_A}=\Gamma, \sum_b\Gamma_{b|y}=\Gamma,\forall b,y,\nonumber\\
	&\tr(\Gamma_{b|y}H^N_{a,c|x,z})=P(a,b,c|x,y,z),\forall a,b,c,x,y,z,\nonumber\\
	&\sum_{a,b,c}P(a,b,c|x,z)=1,
\end{align}
\noindent where $\vec{\Gamma}$ denotes the tuple of matrices $(\Gamma, (\Gamma_{b|y})_{y,b})$.

For $N=2$, we ran the corresponding SDP with the MATLAB packages YALMIP \cite{yalmip} and Mosek \cite{mosek} and obtained the upper bound shown in eq. (\ref{RQT}) in main text.

\section{Wave-plate angles in Alice, Bob and Charlie's measurements}

The wave-plate angles in Alice, Bob and Charlie's measurements are shown in Table S1 and S2, respectively.

\noindent {\bf Table S1.}\\
Bob sets wave-plates to the proper angles to perform measurement $\hat{\Pi}_y$ and $1-\hat{\Pi}_y$ according to the random input $y=y_2 y_1$, respectively. The angle is $0^o$ when the optical axis of the half wave-plate ($H_2$, $H_3$) is oriented vertically in the experiment.
\begin{table}[htbp!]
	\label{tab:PPBSM}
	\begin{tabular}{|c|c|c|c|c|}
		\hline
		{$y=y_2y_1$ (binary)} & H$_2$ & H$_3$ & $\hat{\Pi}_y, b=+1$ & {$1-\hat{\Pi}_y$, $b=-1$}  \\
		\hline
		0 (00) & $@0\degree$ & $@22.5\degree$ & $\ket{\Phi^+}\bra{\Phi^+}$ & $1-\ket{\Phi^+}\bra{\Phi^+}$\\
		\hline
		1 (01) & $@0\degree$ & $@67.5\degree$ & $\ket{\Phi^-}\bra{\Phi^-}$ & $1-\ket{\Phi^-}\bra{\Phi^-}$\\
		\hline
		2 (10) & $@45\degree$ & $@22.5\degree$ & $\ket{\Psi^-}\bra{\Psi^-}$ & $1-\ket{\Psi^-}\bra{\Psi^-}$\\
		\hline
		3 (11) & $@45\degree$ & $@67.5\degree$ & $\ket{\Psi^+}\bra{\Psi^+}$ & $1-\ket{\Psi^+}\bra{\Psi^+}$\\
		\hline
	\end{tabular}
\end{table}

\noindent {\bf Table S2.}\\
Alice and Charlie set the half wave-plates ($H_1$, $H_4$, $H_5$) and quarter wave-plates ($Q_1$) to the proper angles to perform measurements $A_x$ and $C_z$ according to the random inputs $x \in \{0,1,2\}$  and $z \in \{0,1,2,3,4,5\}$, respectively. 
\begin{table}[htbp!]
	\begin{tabular}{|c|c|c|c|}
		\hline
		$x$ & {Q$_1$} & {H$_1$} & $A_x$ \\
		\hline
		0 & $@45\degree$ & $@22.5\degree$ & \makecell[c]{$\sigma_X$}\\
		\hline
		1 & $@45\degree$& $@0\degree$ & $\sigma_Y$\\
		\hline
		2 & $@0\degree$ & $@0\degree$ & $\sigma_Z$\\
		\hline	
		\hline
		{$z$ } &{H$_4$}& {H$_5$} & $C_z$ \\
		\hline
		0 &$@11.25\degree$&$@22.5\degree$ & $\sigma_X+\sigma_Y$\\
		\hline
		1 &$@33.75\degree$&$@22.5\degree$ & $\sigma_X-\sigma_Y$\\
		\hline
		2 &$@22.5\degree$&$@11.25\degree$ & $\sigma_Y+\sigma_Z$\\
		\hline	
		3 &$@22.5\degree$&$@33.75\degree$  & $\sigma_Y-\sigma_Z$\\
		\hline
		4 &$@0\degree$&$@33.75\degree$     & $\sigma_X+\sigma_Z$\\
		\hline
		5 &$@0\degree$&$@11.25\degree$     & $\sigma_X-\sigma_Z$\\
		\hline
	\end{tabular}
	\label{CRM}
\end{table}

\section{Experimental evaluation of the quantity $\sum_{y=0}^{3}p(b=+1|y)$}

We estimate the probability $p(b=+1|y)$ with 
\begin{equation}
	\label{eq:prob}
	\begin{aligned}
		p(b=+1|y)=N_{b=+1|y}/N_{Tot},
	\end{aligned}
\end{equation}
where $N_{Tot}=N_{b=+1|y}+N_{b=-1|y}$ is the number of four-fold coincident events between Alice, Bob, and Charlie, $N_{b=+1|y}$ and $N_{b=-1|y}$ are the number of four-fold coincident events which correspond to Bob's binary outcome $b=+1$ and $b=-1$ with input $y$, respectively. 

Assuming the probability for EPR source 1 (2) to emit a pair of photons is $g_1$ and the probabilities for the pair of photons to be detected by Alice (Charlie) and Bob are respectively $\eta_{A}$ ($\eta_{C}$) and $\eta_{B_1}$ ($\eta_{B_2}$), the probability to detect a four-fold coincidence events is $g_1 g_2 \eta_{A} \eta_{B_1} \eta_{B_2} \eta_{C}$ in one experimental trial. Here $\eta_{B_1}$ and $\eta_{B_2}$ are the overall detection probabilities as the respective photon enters Bob's measurement station.  
Then we have $N_{Tot}=Rg_{1}g_{2}\eta_{A}\eta_{B_1}\eta_{B_2}\eta_{C}$ and 
\begin{equation}
	\begin{aligned}
		p(b=+1|y)=N_{b=+1|y}/N_{Tot}=N_{b=+1|y}/(Rg_{1}g_{2}\eta_{A}\eta_{B_1}\eta_{B_2}\eta_{C}),
	\end{aligned}
\end{equation}
where R is the number of experimental trials. 

We note that the number of two-photon coincidence events detected by Alice and Charlie is given as 
\begin{equation}
	\begin{aligned}
		N_{AC}=Rg_{1}g_{2}\eta_{A}\eta_{C}.
	\end{aligned}
\end{equation}
Similarly, we have
\begin{equation}
	\begin{aligned}
		N_{AB}=Rg_{1}\eta_{A}\eta_{B_1},
		N_{BC}=Rg_{2}\eta_{C}\eta_{B_2},
		N_{A}=Rg_{1}\eta_{A},
		N_{C}=Rg_{2}\eta_{C},
	\end{aligned}
\end{equation}
where $N_{AB}$ and $N_{BC}$ are the two-photon coincidence events between Alice (Charlie) and Bob, $N_A$ and $N_C$ are the numbers of single photon detection events by Alice and Charlie, respectively. 

Then we rewrite $p(b=+1|y)$ as 
\begin{equation}
	\label{eq:probfinal}
	\begin{aligned}
		p(b=+1|y)=\frac{N_{b=+1|y}N_A N_C} {N_{AC} N_{AB} N_{BC}}.
	\end{aligned}
\end{equation} 
We obtain $\sum_{y=0}^{3}p(b=+1|y)=1.0038 \pm 0.0061$ in this experiment.

\section{Quantum state tomography measurements} 

\begin{figure}
	\centering
	\includegraphics[width=\columnwidth]{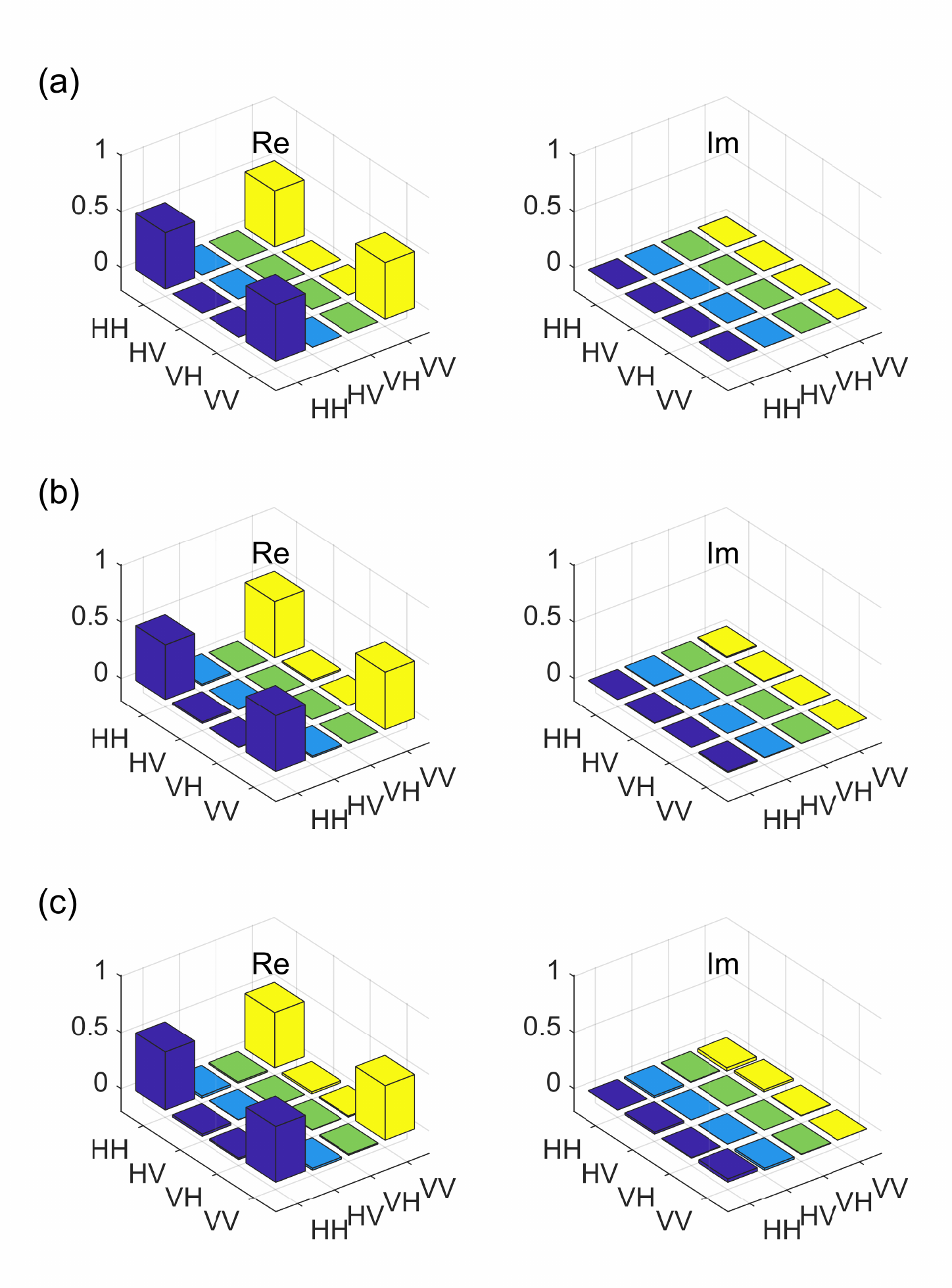}
	\caption{(a) The density matrix of ideal EPR state $\ket{\Phi^+}$. (b)(c) The experimentally reconstructed density matrices of states emitted by EPR sources $S_1$ and $S_2$ with fidelities $0.9955\pm 0.0005$ and $0.9909\pm 0.0007$, respectively.}
	\label{fig:CP-fs1}
\end{figure}

We reconstruct the density matrices of the states emitted by EPR sources $S_1$ and $S_2$ based on quantum state tomography measurement respectively with fidelities $0.9955\pm 0.0005$ and $0.9909\pm 0.0007$ with respect to the ideal state $\ket{\Phi^+}$, as shown in Fig. \ref{fig:CP-fs1}. The uncertainties are obtained using a Monte Carlo routine assuming Poissonian statistics. We also reconstruct the density matrices of the states shared by Alice and Charlie after entanglement swapping measurement by Bob, with fidelities $0.9844\pm 0.0061$, $0.9629\pm 0.0106$, $0.9741\pm 0.0085$ and $0.9661\pm 0.0081$ with respect to ideal states $\ket{\Phi^+}$, $\ket{\Phi^-}$, $\ket{\Psi^+}$, $\ket{\Psi^-}$, respectively, as shown in Fig. \ref{fig:CP-fs1}.  We attribute the imperfection to the residual distinguishability between photons from separate EPR sources. We stress that there is no coherence between pulses of the same laser and of different lasers. 

\begin{figure*}
	\centering
	\includegraphics[width=0.8\linewidth]{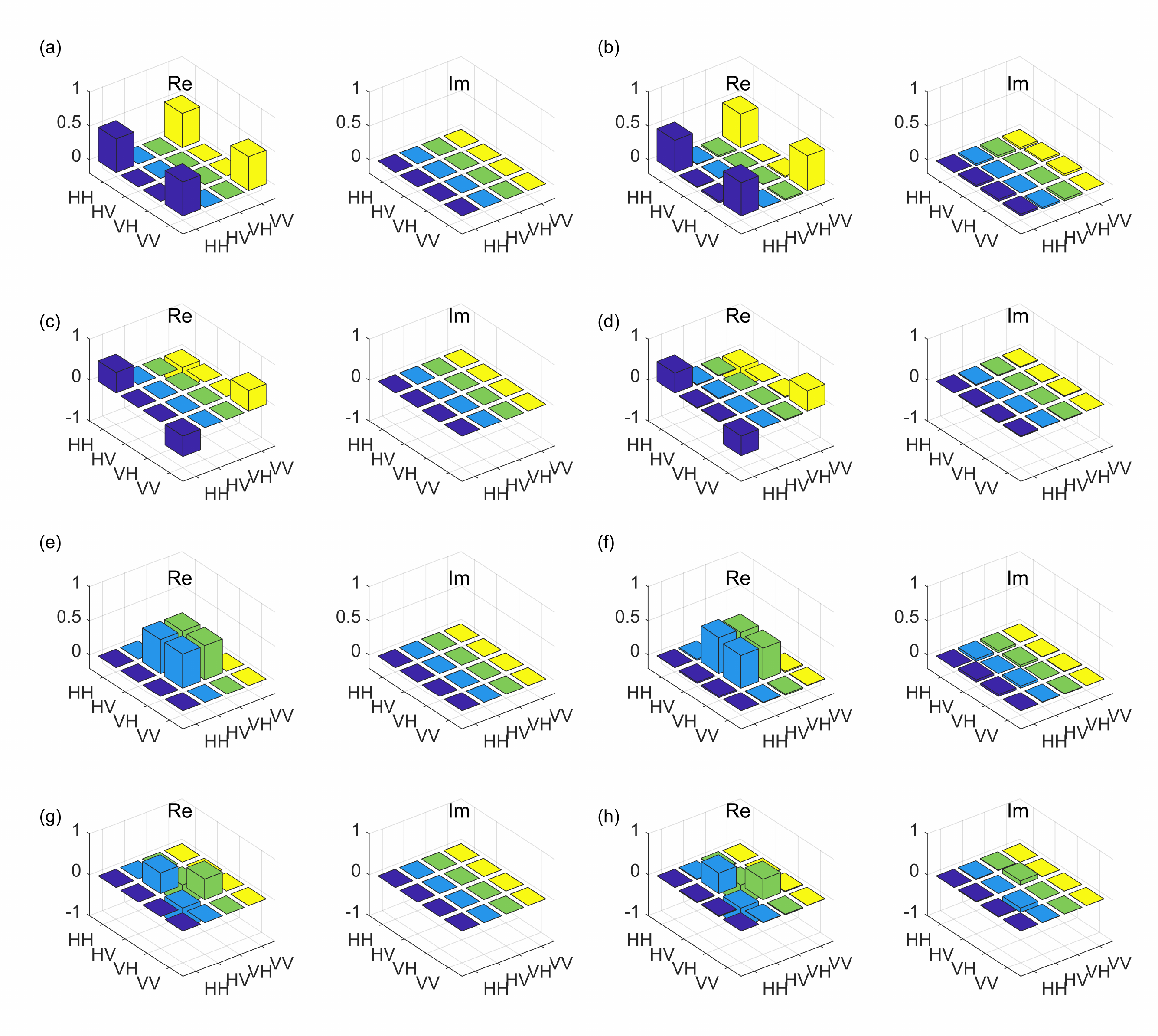}
	\caption{(a)(c)(e)(g) Density matrices of ideal EPR states $\ket{\Phi^+}$, $\ket{\Phi^-}$, $\ket{\Psi^+}$, $\ket{\Psi^-}$. (b)(d)(f)(h) Reconstructed density matrices of states shared by Alice and Charlie in the experiment.}
	\label{fig:CP-fs2}
\end{figure*}

\begin{figure}
	\centering
	\includegraphics[width=\columnwidth]{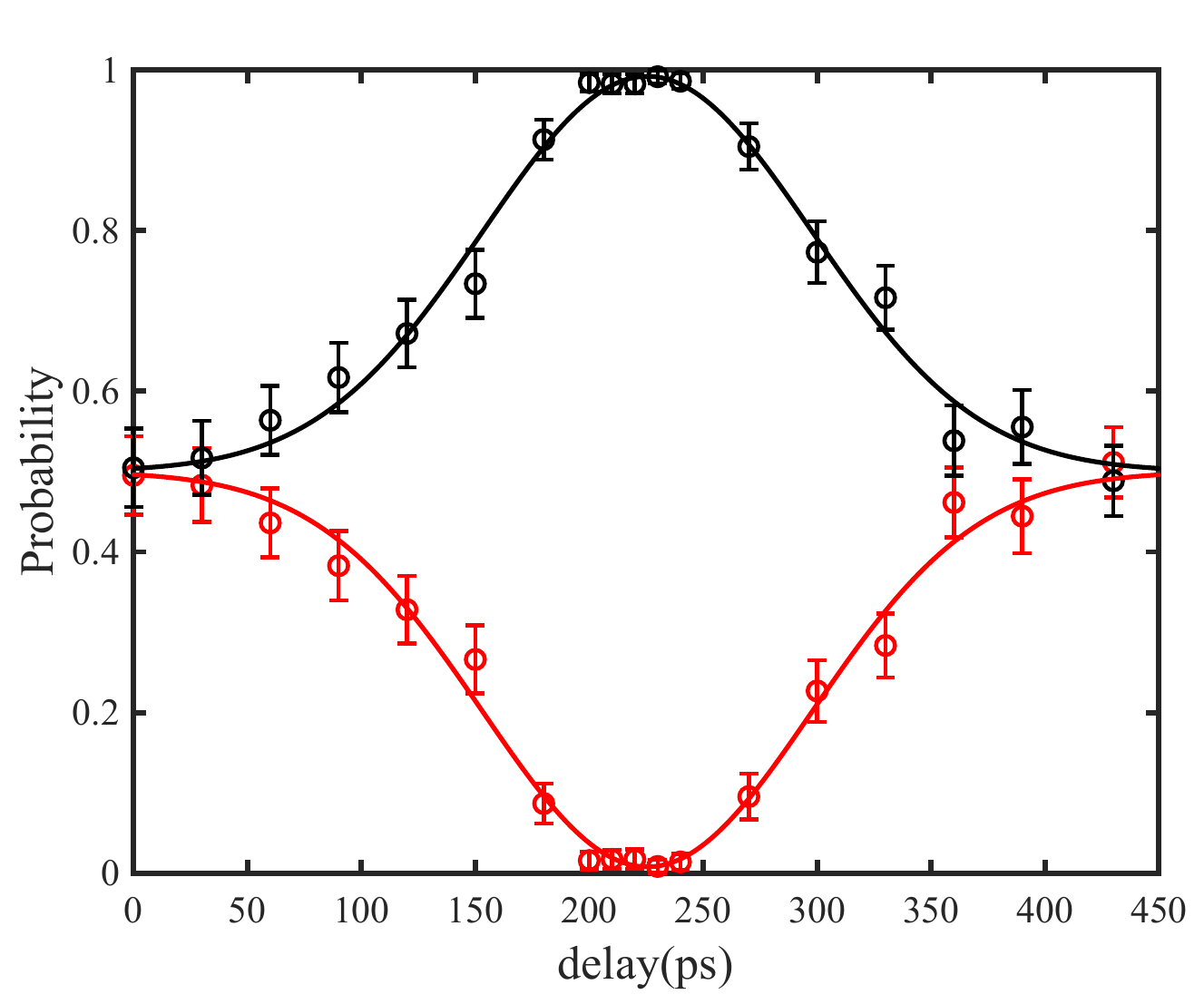}
	\caption{Two-photon Hong-Ou-Mandel interference with a fitted visibility $V_I=$0.9844 in our experiment.}
	\label{fig:HOM}
\end{figure}

\section{Complex quantum theory emulation of the experiment}

We emulate the experiment based on complex quantum theory, for which we consider experimental imperfections.
We model the state of a pair of photons emitted by the EPR source by a density operator $\rho_{EPR}=v_{E}\ket{\Phi^+}\bra{\Phi^+}+(1-v_{E})I/4$, and the four Bell state measurements performed by Bob by $M_{0}=v_{I}\ket{\Phi^+}\bra{\Phi^+}+(1-v_{I})(\ket{\Phi^+}\bra{\Phi^+}+\ket{\Phi^-}\bra{\Phi^-})/2$, $M_{1}=v_{I}\ket{\Phi^-}\bra{\Phi^-}+(1-v_{I})(\ket{\Phi^+}\bra{\Phi^+}+\ket{\Phi^-}\bra{\Phi^-})/2$, $M_{2}=v_{I}\ket{\Psi^-}\bra{\Psi^-}+(1-v_{I})(\ket{\Psi^+}\bra{\Psi^+}+\ket{\Psi^-}\bra{\Psi^-})/2$, $M_{3}=v_{I}\ket{\Psi^+}\bra{\Psi^+}+(1-v_{I})(\ket{\Psi^+}\bra{\Psi^+}+\ket{\Psi^-}\bra{\Psi^-})/2$, where $v_I$ is the visibility of the Hong-Ou-Mandel interference with two photons from EPR sources $S_1$ and $S_2$ (Fig. \ref{fig:HOM}). We then evaluate the correlation functions and the value of $W$ for different values of $v_{E}$ and $v_{I}$. 
The fitted visibility is $V_I=0.9844$ in our experiment. 
\end{appendix}

\end{document}